\documentclass[conference]{IEEEtran}
\IEEEoverridecommandlockouts
\usepackage{cite}
\usepackage{amsmath,amssymb,amsfonts}
\usepackage{algorithmic}
\usepackage{graphicx}
\usepackage{textcomp}
\usepackage{xcolor}
\usepackage{balance}

\def\BibTeX{{\rm B\kern-.05em{\sc i\kern-.025em b}\kern-.08em
		T\kern-.1667em\lower.7ex\hbox{E}\kern-.125emX}}

\makeatletter
\def\endthebibliography{%
	\def\@noitemerr{\@latex@warning{Empty `thebibliography' environment}}%
	\endlist
}
\makeatother

\begin{document}
	
	\title{Pathloss modeling for in-body optical wireless communications}
	
	\author{\IEEEauthorblockN{Stylianos E. Trevlakis,
			Alexandros-Apostolos A. Boulogeorgos, and
			Nestor D. Chatzidiamantis}
		\IEEEauthorblockA{Department of Electrical and Computer Engineering, 
			Aristotle University of Thessaloniki, 
			Thessaloniki, 
			Greece, 54124\\
			Emails: $\lbrace$trevlakis; nestoras$\rbrace$@auth.gr; al.boulogeorgos@ieee.org}}
	
	\maketitle
	
	\begin{abstract}
		Optical wireless communications (OWCs) have been recognized as a candidate enabler of next generation in-body nano-scale networks and implants. The development of an accurate channel model capable of accommodating the particularities of different type of tissues is expected to boost the design of optimized communication protocols for such applications. Motivated by this, this paper focuses on presenting a general pathloss model for in-body OWCs. In particular, we use experimental measurements in order to extract analytical expressions for the absorption coefficients of the five main tissues' constitutions, namely oxygenated and de-oxygenated blood, water, fat, and melanin. Building upon these expressions, we derive a general formula for the absorption coefficient evaluation of any biological tissue. To verify the validity of this formula, we compute the absorption coefficient of complex tissues and compare them against respective experimental results reported by independent research works. Interestingly, we observe that the analytical formula has high accuracy and is capable of modeling the pathloss and, therefore, the penetration depth in complex tissues.
	\end{abstract}
	
	
	\section{Introduction}
	Optical wireless communication (OWC) based in-body biomedical applications have attracted a significant amount of attention over the last couple of years, due to the performance excellency (in terms of reliability, speed, energy efficiency and latency) that they are expected to achieve~\cite{Trevlakis2018mocast,Trevlakis2018mdpi,Trevlakis2018spawc,Trevlakis2019owci,Trevlakis2020aoci,trevlakis2020commag}. In order to optimize the in-body OWC system performance, an accurate channel model that takes into account the tissue characteristics needs to be employed. 
	
	Scanning the technical literature, it can be observed that most efforts focus on quantifying the optical characteristics of specific tissues at certain wavelengths~\cite{Tseng2011,Salomatina2006,Shimojo2020,Sandell2011,Pifferi2004,Spinelli2004,Bashkatov2006,Ugryumova2004,Zhao2005,Yaroslavsky2002,Zee1993}. Also, in~\cite{Tseng2011,Salomatina2006,Shimojo2020}, the authors performed measurements for the optical properties of both healthy and cancerous skin in the visible and near-infrared spectral range. In~\cite{Sandell2011,Pifferi2004,Spinelli2004}, experiments were performed that quantified the optical properties of human female breast tissues in multiple wavelengths and over different distances. Furthermore, the authors  in~\cite{Bashkatov2006,Ugryumova2004} evaluated the light absorption and scattering of bone tissue for various wavelengths in the visible and near-infrared spectrum. Finally, in~\cite{Zhao2005,Yaroslavsky2002,Zee1993}, the optical properties of human brain tissue at various ages were studied in the visible spectrum. However, such results are not always useful for other researchers due to various reasons. Firstly, they may not include the required wavelengths of interest. Secondly, even if the wavelength is available, the constitution of a tissue is different enough between distinct individuals that the results cannot be regarded as confident. As a result, the need arises for the development of method that estimates the optical properties of a tissue based on its constitution.
	
	To this end, specific formulas have been reported for the pathloss evaluation of a generic tissue that take into account the variable amounts of its constituents (i.e. blood, water, fat, melanin), but require their optical properties at the exact transmission wavelength, which hinders the use of these formulas~\cite{Jacques2013,TrevlakisOvsE}. Motivated by this, this paper derives a novel mathematical model, which requires no experimental measurements for the calculation of the pathloss for in-body OWCs. Based on the aforementioned, the technical contribution of the this work is summarized as follows:
	\begin{itemize}
		\item We extract analytical expressions for the absorption coefficients of the five main tissues' constitutions, namely oxygenated and de-oxygenated blood, water, fat, and melanin.	
		\item Based on the these expressions, we derive a general formula for the absorption coefficient evaluation of any biological tissue.
		\item We present the analytical results for the absorption coefficients of complex human tissues, such as brain, bone, breast and skin, and compare them with experimental results reported by independent works that prove the validity of the presented mathematical framework.
		\item Finally, we illustrate the pathloss as a function of the transmission wavelength for different complex tissues and tissue thickness, and provide discussions that highlight useful insights for the design of communication protocols.
	\end{itemize}
	
	The rest of this paper is organized as follows. Section~\ref{S:pathloss_model} is devoted in presenting the pathloss model based on the absorption properties of the constituents of any generic tissue. Section~\ref{S:results} presents respective numerical results that verify the mathematical framework and insightful discussions, which highlight design guidelines for communication protocols. Finally, closing remarks are summarized in Section~\ref{S:conclusion}.
	
	\textit{Notations:} Unless stated otherwise, in this paper, $\exp(\cdot)$ represents the exponential function, while $\cos(\cdot)$ and $\sin(\cdot)$ stand for the cosine and sine functions, respectively.  
	
	\section{Pathloss Model} \label{S:pathloss_model}
	The pathloss caused by the absorption of biological tissue can be modeled as
	\begin{align}
		L = \exp\left(\mu_a \delta\right) ,
	\end{align}
	where $\delta$ is the transmission distance and $\mu_a$ is the absorption coefficient, which can be defined as
	\begin{align} \label{eq:mu_1}
		\mu_a = -\frac{1}{T} \frac{\partial T}{\partial \delta} ,
	\end{align}
	with $T$ denoting the fraction of residual optical radiation at $\delta$. Thus, the fractional change of the intensity of the incident light can be obtained as
	\begin{align}
		T = \exp\left(-\mu_a \delta\right) .
	\end{align}
	
	The absorption coefficient can be expressed as the sum of all the tissue's constituents, namely water, melanin, fat, oxygenated blood and de-oxygenated blood. Thus,~\eqref{eq:mu_1} can be analytically expressed as in~\cite{Jacques2013}
	\begin{align} \label{eq:mu_a}
		\begin{split}
			\mu_a = &B S {\mu_a}_{(oBl)} + B \left(1-S\right) {\mu_a}_{(dBl)} \\
			&+ W {\mu_a}_{(w)} + F {\mu_a}_{(f)} + M {\mu_a}_{(m)} 
		\end{split} ,
	\end{align}
	where ${\mu_a}_{(i)}$ represents the absorption coefficient of the $i$-th constituent, while $B$, $W$, $F$, and $M$ represent the blood, water, fat, and melanin volume fractions, respectively. Finally, $S$ denotes the oxygen saturation of hemoglobin. 
	
	From~\eqref{eq:mu_a}, it becomes evident that, in order to evaluate the absorption coefficients and volume fractions of each constituent in order to calculate the complete absorption coefficient of any tissue. Although the techniques for measuring optical properties evolve in terms of accuracy and speed, the fact that tissue’s optical properties must be regarded as variables between different tissues, people and even times, hinders their mathematical modeling. These variations, on the one hand are inherent on the individuality of each person, while on the other hand they are subject to the measurement techniques and tissue preparation protocols. However, the absorption coefficients of the tissue's constituents have been measured in existing literature and are mainly dependent on the wavelength of the transmitted optical radiation.
	
	\begin{figure}
		\centering\includegraphics[width=0.8\linewidth,trim=0 0 0 0,clip=false]{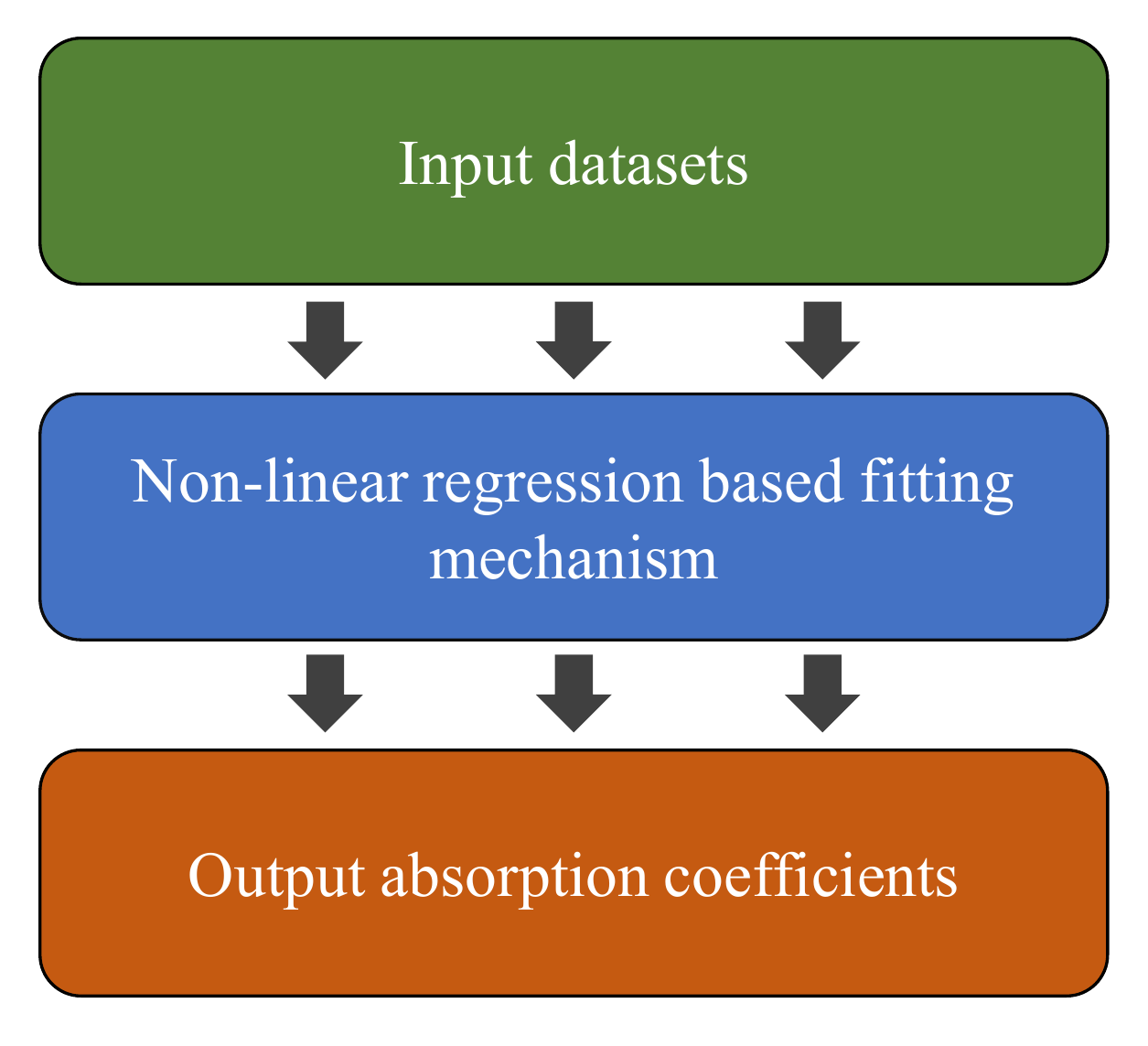}
		\caption{Machine learning based fitting mechanism.}\label{fig:fitting}
	\end{figure}
	In this direction, we used the machine learning mechanism based on non-linear regression, which is presented in~Fig.\ref{fig:fitting}, to extract analytical expressions for the absorption coefficients of oxygenated and de-oxygenated blood, water, fat, and melanin from on the experimental datasets for each of the constituents as input~\cite{boulogeorgos2020MLinNanoBio}. In particular, absorption coefficient datasets of oxygenated and de-oxygenated blood have been provided in~\cite{Takatani1979,Moaveni1970,Schmitt1986,Zhao2017}. The analytical expressions of the absorption coefficients of oxygenated and de-oxygenated blood, have been fitted on the experimental results using the sum of Gaussian functions and can be expressed as 
	\begin{align} \label{eq:m_dBlood}
		{\mu_a}_{(dBl)}\left(\lambda\right) = \sum_{i=1}^{4} {a_i}_{(dBl)} \exp\left(-\left(\frac{\lambda-{b_i}_{(dBl)}}{{c_i}_{(dBl)}}\right)^2\right) ,	
	\end{align}
	and 
	\begin{align} \label{eq:m_oBlood}
		{\mu_a}_{(oBl)}\left(\lambda\right) = \sum_{i=1}^{5} {a_i}_{(oBl)} \exp\left(-\left(\frac{\lambda-{b_i}_{(oBl)}}{{c_i}_{(oBl)}}\right)^2\right) .
	\end{align} 
	Furthermore, the absorption coefficient of water has been evaluated in several contributions~\cite{Hale1973,Zolotarev1969,Segelstein1981}, which were used to extract a Fourier series that accurately describes the absorption coefficient as a function of the transmission wavelength. The extracted analytical expression of the absorption coefficient of water can be written as
	\begin{align} \label{eq:m_water}		
		{\mu_a}_{(w)}\left(\lambda\right) = {a_0}_{(w)} + \sum_{i=1}^{7} {a_i}_{(w)} \cos(i w \lambda) + {b_i}_{(w)} \sin(i w \lambda) .
	\end{align}
	Moreover, the absorption of fat is highly dependent on the origin of the fat. For proper measurement of the optical properties, the fat tissue must be purified and dehydrated. This necessary preparation may cause inconsistencies between different published works. However, the results presented in~\cite{Bashkatov2005} coincide with other works in the visible spectrum and, thus, they are selected for extracting the mathematical model for the absorption coefficient of fat. The analytical expression was extracted by fitting the experimental measurements with a sum of Gaussian functions and is given by
	\begin{align} \label{eq:m_fat}
		{\mu_a}_{(f)}\left(\lambda\right) = \sum_{i=1}^{5} {a_i}_{(f)} \exp\left(-\left(\frac{\lambda-{b_i}_{(f)}}{{c_i}_{(f)}}\right)^2\right) .
	\end{align}
	It should be highlighted that the parameters of the previously extracted analytical expressions are provided in Table~\ref{tbl:parameters}. Finally, the experimental data available in open literature for the absorption coefficient of melanin are highly consistent for the visible spectrum~\cite{Jacques1991,Zonios2008,Jacques2013}. Based on these results, the analytical expression of the absorption coefficient of melanin is given as
	\begin{align} \label{eq:m_melanin}
		{\mu_a}_{(m)}\left(\lambda\right) = {\mu_a}_{(m)}\left(550\right) \left(\frac{\lambda}{550}\right)^{-3} ,
	\end{align}
	where ${\mu_a}_{(m)}\left(550\right)$ denotes the absorption coefficient of melanin at $550\text{ }\mathrm{nm}$, which is equal to $519\text{ }\mathrm{cm}^{-1}$.
	
	\begin{table} 
		\centering
		\caption{Fitting parameters for constituent's absorption coefficient.}
		\begin{tabular}{|c|c|c|c|c|}
			\hline
			& dBlood & oBlood & water & fat \\
			\hline
			$a_0$ & - & - & $ 324.1 $ & - \\
			\hline
			$a_1$ & $ 38.63 $ & $ 14 $ & $ 102.2 $ & $ 33.53 $ \\
			\hline
			$a_2$ & $ 60.18 $ & $ 13.75 $ & $ -568 $ & $ 50.09 $ \\
			\hline
			$a_3$ & $ 25.11 $ & $ 29.69 $ & $ -126.6 $ & $ 3.66 $ \\
			\hline
			$a_4$ & $ 2.988 $ & $4.317 \times 10^{15}$ & $ 236.8 $ & $ 2.5 $ \\
			\hline
			$a_5$ & - & $ -34.3 $ & $ 73 $ & $ 19.86 $ \\
			\hline
			$a_6$ & - & - & $ -40.53 $ & - \\
			\hline
			$a_7$ & - & - & $ -12.92 $ & - \\
			\hline
			$b_1$ & $ 423.9 $ & $ 419.7 $ & $ 697.9 $ & $ 411.5 $ \\
			\hline
			$b_2$ & $ 31.57 $ & $ 581.5 $ & $ 121.7 $ & $ 968.7 $ \\
			\hline
			$b_3$ & $ 559.3 $ & $ 559.9 $ & $ -395.3 $ & $ 742.9 $ \\
			\hline
			$b_4$ & $ 664.7 $ & $ -25880 $ & $ -107.1 $ & $ 671.2 $ \\
			\hline
			$b_5$ & - & $ 642.6 $ & $ 115.6 $ & $ 513.8 $ \\
			\hline
			$b_6$ & - & - & $ 35.46 $ & - \\
			\hline
			$b_7$ & - & - & -$ 8.373 $ & - \\
			\hline
			$c_1$ & $ 33.06 $ & $ 16.97 $ & - & $ 38.38 $ \\
			\hline
			$c_2$ & $ 660.8 $ & $ 11.68 $ & - & $ 525.9 $ \\
			\hline
			$c_3$ & $ 59.08 $ & $ 46.71 $ & - & $ 80.22 $ \\
			\hline
			$c_4$ & $ 28.53 $ & $ 4668 $ & - & $ 32.97 $ \\
			\hline
			$c_5$ & - & $ 162.5 $ & - & $ 119.2 $ \\
			\hline
			$w$ & - & - & $ 0.006663 $ & - \\
			\hline
		\end{tabular}
		\label{tbl:parameters}
	\end{table}
	
	\section{Results \& Discussion} \label{S:results}
	This section focuses not only on the verification of the mathematical framework presented in Section~\ref{S:pathloss_model} via comparing the experimental data with the analytical expressions derived from it, but also on the illustration of its accuracy in modeling complex biological tissues, such as skin, bone, breast, and brain tissue. Finally, the pathloss is evaluated for each complex tissue and insightful discussions are provided that illustrates important design guidelines for communication protocols.
	
\begin{figure}
		\centering\includegraphics[width=0.95\linewidth,trim=0 0 0 0,clip=false]{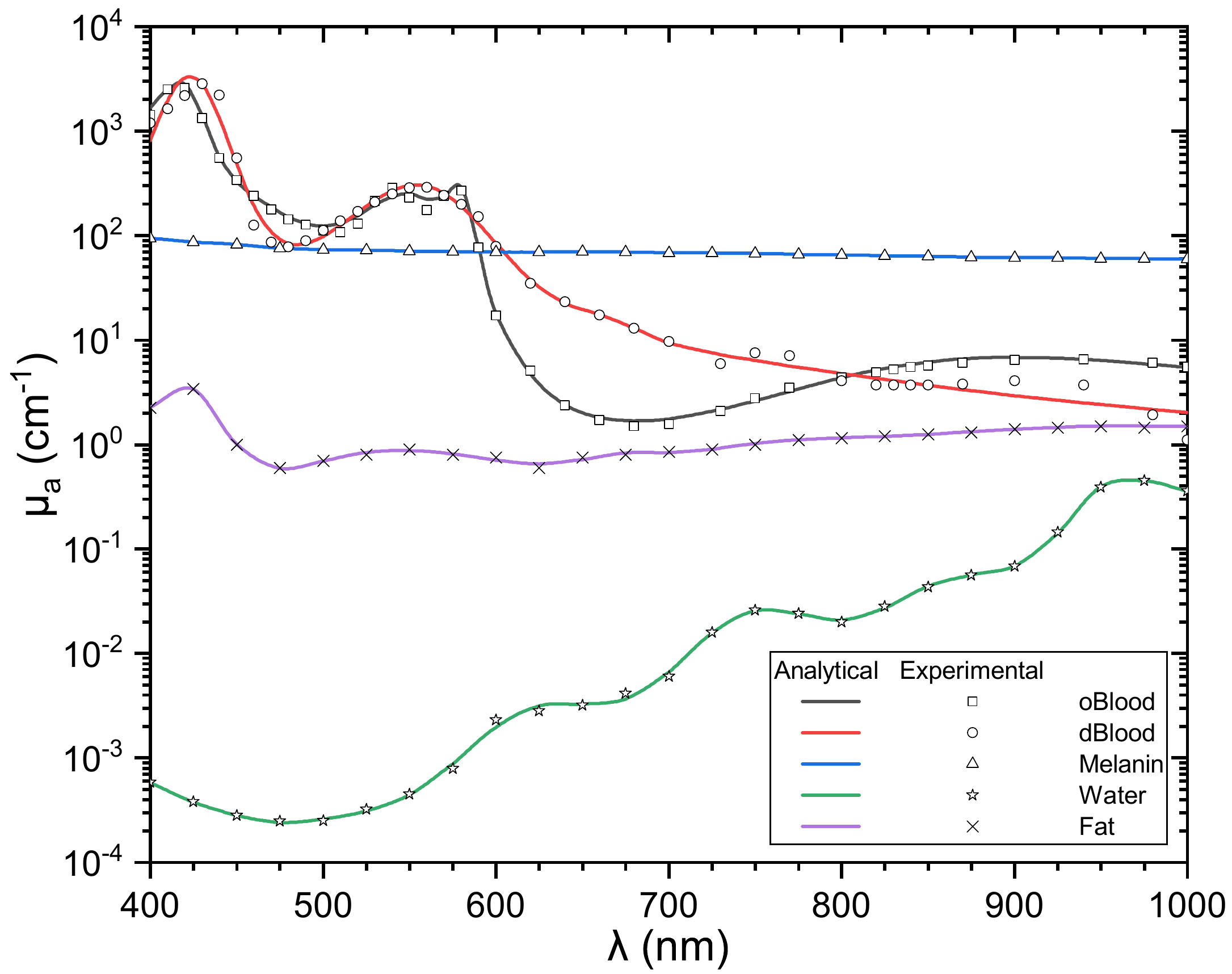}
		\caption{Absorption coefficients of generic tissue constituents as a function of the transmission wavelength.}\label{fig:complete_fitting}
	\end{figure}

	Fig.~\ref{fig:complete_fitting} depicts the experimental data for each constituent's absorption coefficient against the analytical results extracted from~\eqref{eq:m_oBlood} through~\eqref{eq:m_melanin}. In more detail, the analytical expressions for the oxygenated and de-oxygenated blood, water, melanin, and fat are drawn in black, red, green, blue, and purple color, respectively, while the corresponding experimental data are represented by square, circle, star, triangle, and cross symbols. The analytical and experimental results coincide, which proves the validity of the analytical expressions. Another interesting observation from this figure is that the absorption coefficient of blood is higher than the rest constituents between $400$ and $600\text{ }\mathrm{nm}$, while it is still among the most influential for higher wavelength values. This highlights that even with relatively low volume fraction, blood plays an important role in the total absorption coefficient of any generic tissue. Furthermore, it is evident that as $\lambda$ increases, the absorption coefficient increases as well, which highlights the importance of carefully selecting the transmission wavelength used in tissues with high concentration in water. Moreover, we observe that the absorption coefficient of fat receives values around $1\text{ }\mathrm{cm}^{-1}$ with very small variations, which constitutes its impact on the total absorption of generic tissues stable throughout the visible spectrum. Finally, it becomes evident that the absorption of melanin is among the highest between the generic tissue constituents, but the most consistent throughout the visible spectrum. This illustrates the importance of the concentration of melanin in the tissue under investigation and, at the same time, the negligible effect of the transmission wavelength on the absorption due to melanin. 
	
	It should be highlighted that, the volume fraction of any of the constituents plays a very important role in the final form of the absorption coefficient. For example, if a tissue is rich in water, the impact of the absorption coefficient of water after it is multiplied by the water volume fraction can affect the total absorption coefficient significantly, even if the absorption coefficient of water seems insignificant on its own. On the other hand, the impact of a constituent with high absorption coefficient, such as melanin, can be diminished if it has a low volume fraction. As a result, although the absorption coefficients presented in~Fig.~\ref{fig:complete_fitting} are very useful for determining which constituents can affect the total absorption coefficient of the generic tissue, it is not an absolute metric and must be used with caution for making assumptions. In the rest of this section, we present experimental results from the open literature for the absorption coefficients of complex human tissues and compare them to the estimation calculated based on the mathematical framework that is presented above.
	
	\begin{table}
		\centering
		\addtolength{\tabcolsep}{-2pt}
		\caption{Tissue parameters related to optical absorption for skin, bone, brain and breast tissue.}
		\begin{tabular}{|c|c|c|c|c|c|c|}
			\hline
			Tissue & $B(\%)$ & $S(\%)$ & $W(\%)$ & $F(\%)$ & $M(\%)$ & Source \\
			\hline
			Skin & $0.41$ & $99.2$ & $26.1$ & $22.5$ & $1.15$ & \cite{Tseng2011,Salomatina2006,Sandell2011,Shimojo2020} \\
			\hline
			Breast & $0.5$ & $52$ & $50$ & $13$ & $0$ & \cite{Pifferi2004,Sandell2011,Spinelli2004} \\
			\hline
			Bone & $0.15$ & $30$ & $30$ & $7$ & $0$ & \cite{Sandell2011,Bashkatov2006,Ugryumova2004} \\
			\hline
			Brain & $1.71$ & $58.7$ & $50$ & $20$ & $0$ & \cite{Zhao2005,Yaroslavsky2002,Zee1993} \\
			\hline
		\end{tabular}
		\label{tbl:tissue_properties}
	\end{table}
	
	\begin{figure}
		\centering\includegraphics[width=0.95\linewidth,trim=0 0 0 0,clip=false]{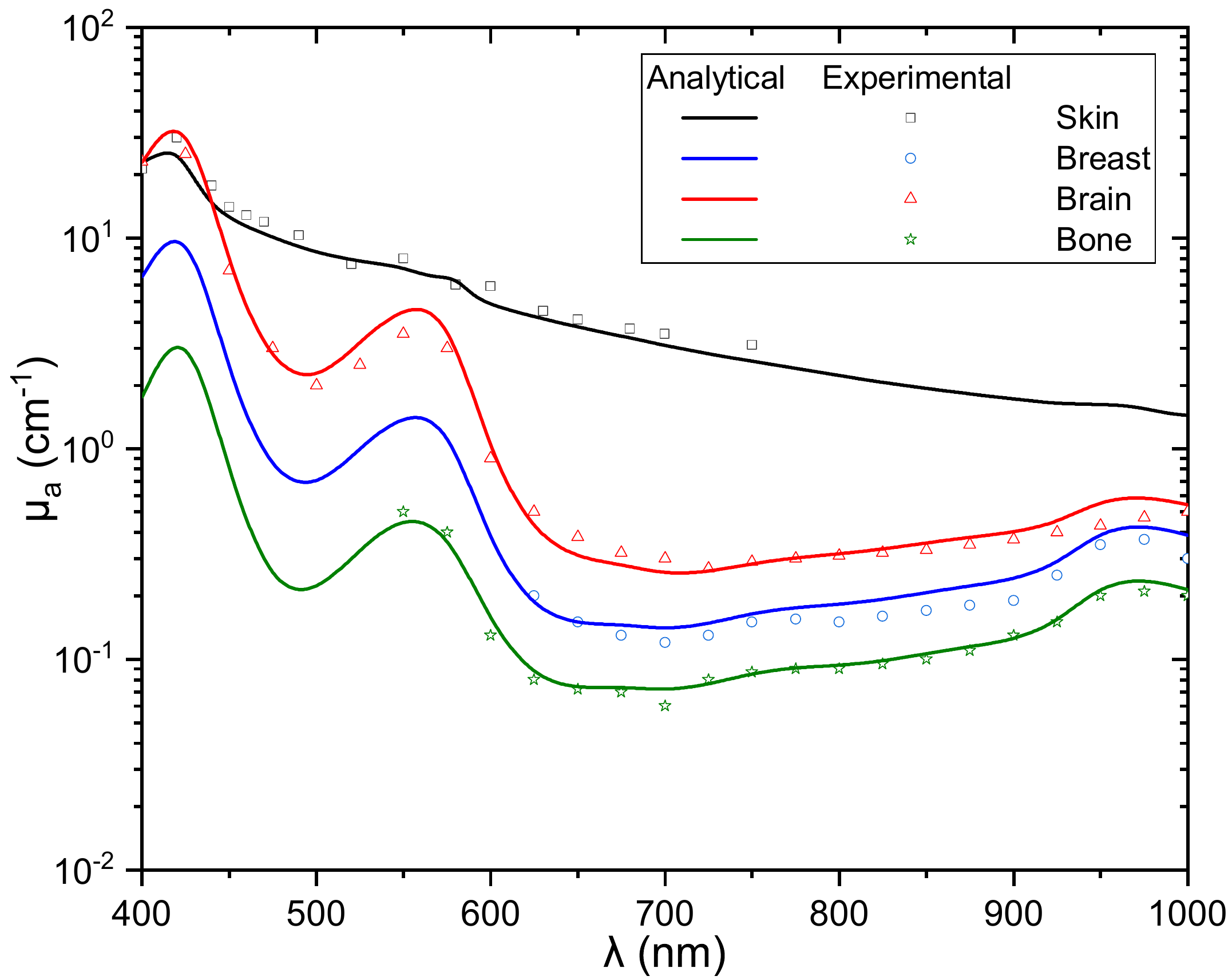}
		\caption{Total absorption coefficient of complex tissues as a function of the transmission wavelength.}\label{fig:complex_tissues_absorption_coefficient}
	\end{figure}
	In Fig.~\ref{fig:complex_tissues_absorption_coefficient}, the absorption coefficients of complex tissues, such as skin, bone, brain and breast, are presented as a function of the transmission wavelength. The analytical expressions and the experimental results are depicted as continuous lines and circles, respectively. The experimental parameters for each tissue are provided in Table~\ref{tbl:tissue_properties} alongside their sources. From this figure, we observe that the analytical expression for the total absorption coefficient provides a very close fit to the experimental data, which verifies the validity of the extracted expressions and provides proof that the presented mathematical framework can describe the optical absorption of generic tissues accurately. Furthermore, it is obvious that the skin absorption coefficient has the most linear behavior out of all the plotted tissues. This happens because of the increased concentration of melanin in the skin, which leads to increased absorption for higher wavelengths. On the other hand, the absorption coefficients of other tissues, which have increased blood and water concentrations, bare a strong resemblance to the blood absorption coefficient in the region between $400$ and $600\text{ }\mathrm{nm}$, while the impact of the absorption coefficient of water becomes visible after $900\text{ }\mathrm{nm}$. This happens because the higher blood and water volume fractions result in increased absorption in the wavelengths where each absorption coefficient has relatively high values.
	
	\begin{figure}
		\centering\includegraphics[width=0.95\linewidth,trim=0 0 0 0,clip=false]{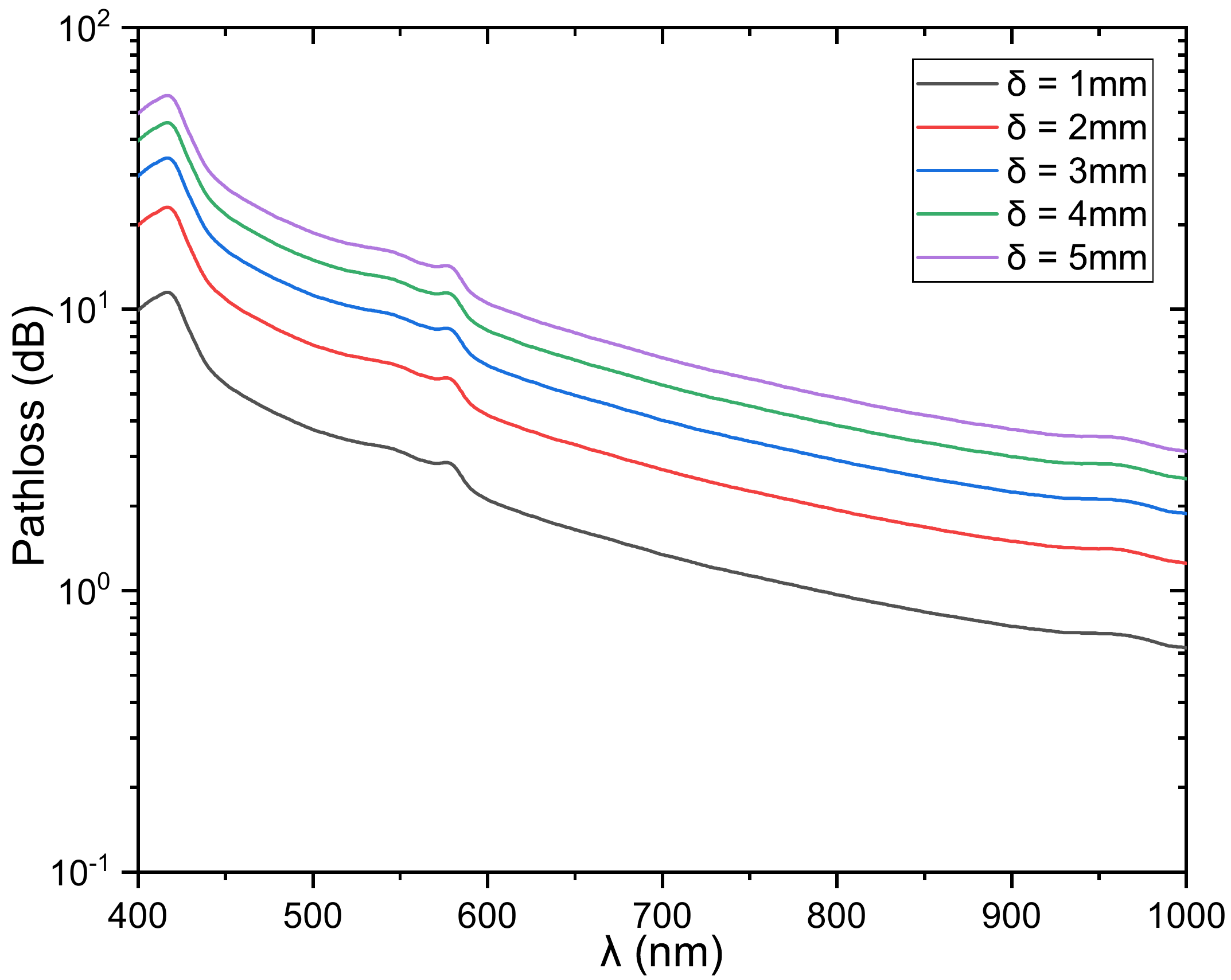}
		\caption{Pathloss due to skin absorption as a function of the transmission wavelength for different values of skin thickness.}\label{fig:skin_pathloss}
	\end{figure}
	Having extracted the accurate absorption coefficients for the various complex tissues, we can calculate the pathloss. To this end, Fig.~\ref{fig:skin_pathloss} depicts the pathloss as a function of the wavelength due to absorption in skin tissues with different values of thickness. By observing this figure, it becomes evident that the pathloss decreases with the wavelength and increases with the skin thickness. Thus, the optimal wavelength in the plotted region is $1000\text{ }\mathrm{nm}$. Furthermore, in the following we assume a transmission window to be the spectrum region where the pathloss does not exceed $6\text{ }\mathrm{dB}$, i.e. the residual optical signal is at least a quarter of the transmitted one. Thus, for $\delta = 1\text{ }\mathrm{mm}$ a transmission window exists for wavelength values higher than $450\text{ }\mathrm{nm}$. However, this transmission windows shrinks as the transmission distance increases. For example, for $\delta = 3\text{ }\mathrm{mm}$ it reduces to wavelengths higher than $650\text{ }\mathrm{nm}$, while for $\delta = 3\text{ }\mathrm{mm}$ it becomes even smaller for wavelengths higher than $650\text{ }\mathrm{nm}$.
	
	\begin{figure}
		\centering\includegraphics[width=0.95\linewidth,trim=0 0 0 0,clip=false]{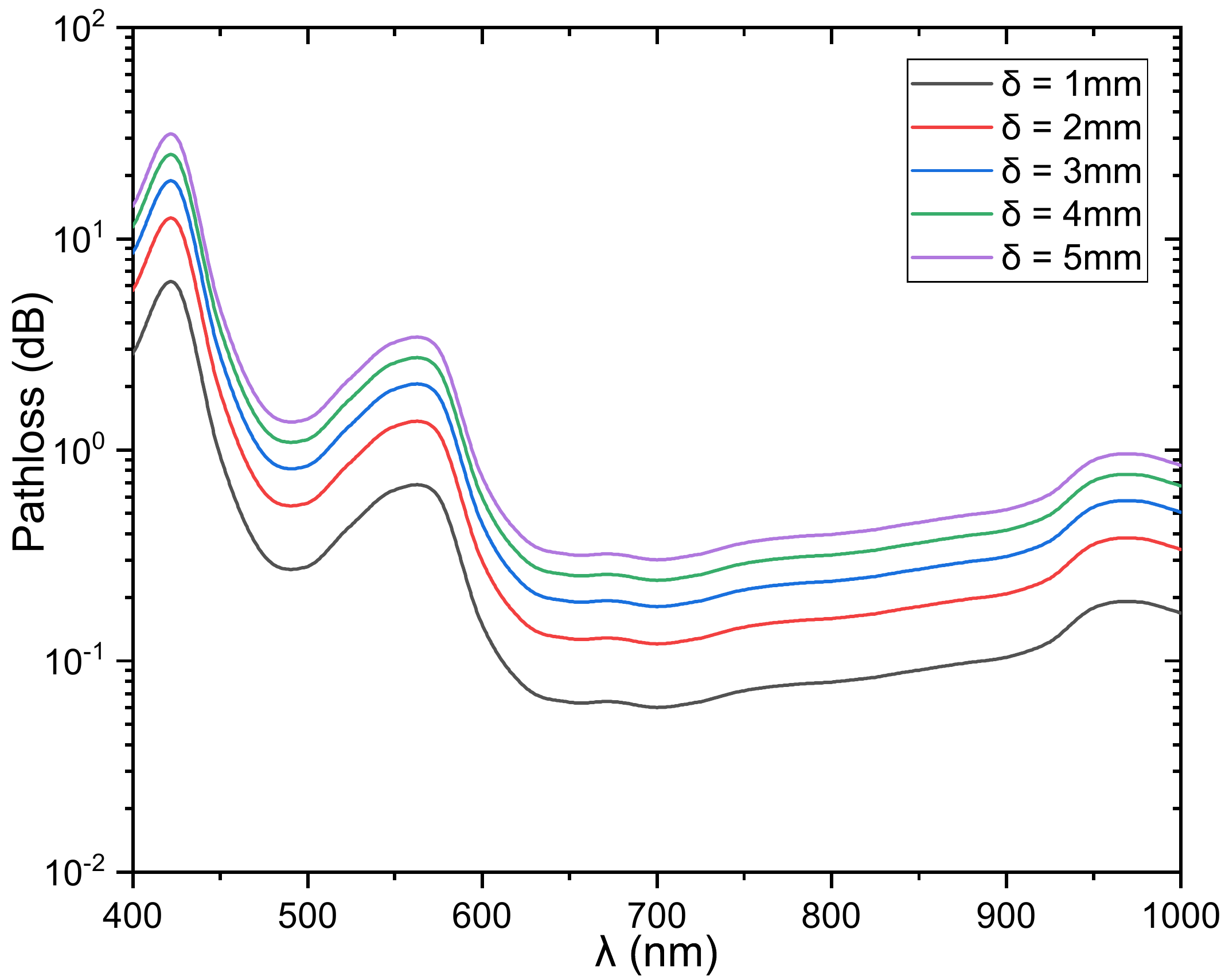}
		\caption{Pathloss due to breast absorption as a function of the transmission wavelength for different values of breast thickness.}\label{fig:breast_pathloss}
	\end{figure}
	In Fig.~\ref{fig:breast_pathloss}, the pathloss of breast tissue is presented with regard to the transmission wavelength for various values of transmission distance. We observe that, as the $\delta$ increases, the pathloss increases as well. On the contrary, the wavelength influences the pathloss in a non linear manner. Moreover, a single transmission window exists for all the plotted values of $\delta$ and is located after $550\text{ }\mathrm{nm}$. However, for higher values of $\delta$ this transmission window will be divided in two. Finally the optimal transmission wavelength for breast tissue is $700\text{ }\mathrm{nm}$.
	
	\begin{figure}
		\centering\includegraphics[width=0.95\linewidth,trim=0 0 0 0,clip=false]{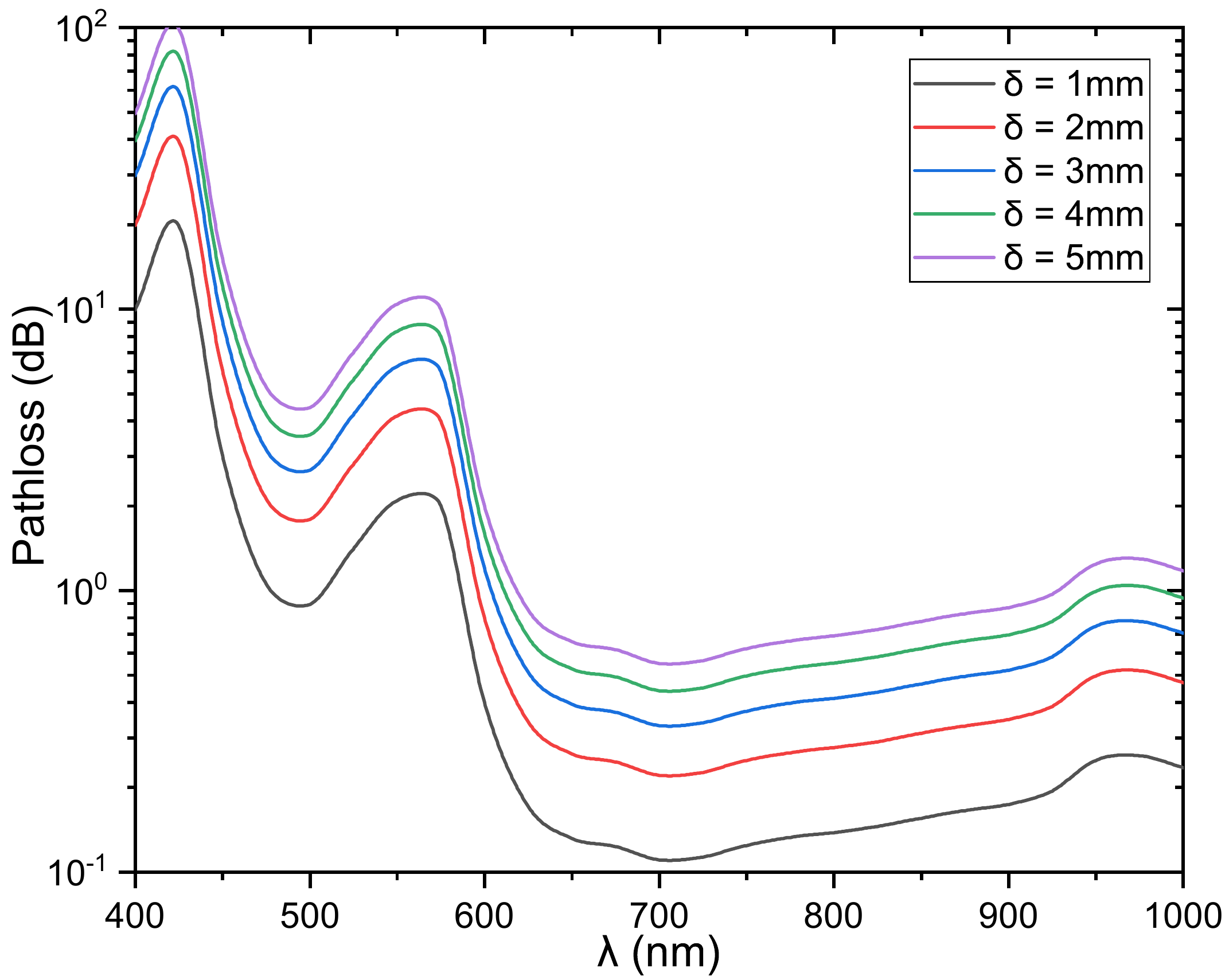}
		\caption{Pathloss due to brain absorption as a function of the transmission wavelength for different values of brain thickness.}\label{fig:brain_pathloss}
	\end{figure}
	Fig.~\ref{fig:brain_pathloss} illustrates the pathloss as a function of the wavelength for different values of tissue thickness. As expected, for higher values of $\delta$ the pathloss is also higher, while the behavior of pathloss for wavelength variations is not linear. For example, as $\lambda$ increases from $500$ to $550\text{ }\mathrm{nm}$ the pathloss increases, while for the same increase from $550$ to $600\text{ }\mathrm{nm}$, pathloss decreases. Also, the optimal transmission wavelength is $700\text{ }\mathrm{nm}$. Furthermore, two transmission windows exist for $\delta = 1\text{ }\mathrm{mm}$. The first is between $450$ and $550\text{ }\mathrm{nm}$, while the second after $600\text{ }\mathrm{nm}$. However, as the transmission distance increases, only the second window will be valid. 
	
	\begin{figure}
		\centering\includegraphics[width=0.95\linewidth,trim=0 0 0 0,clip=false]{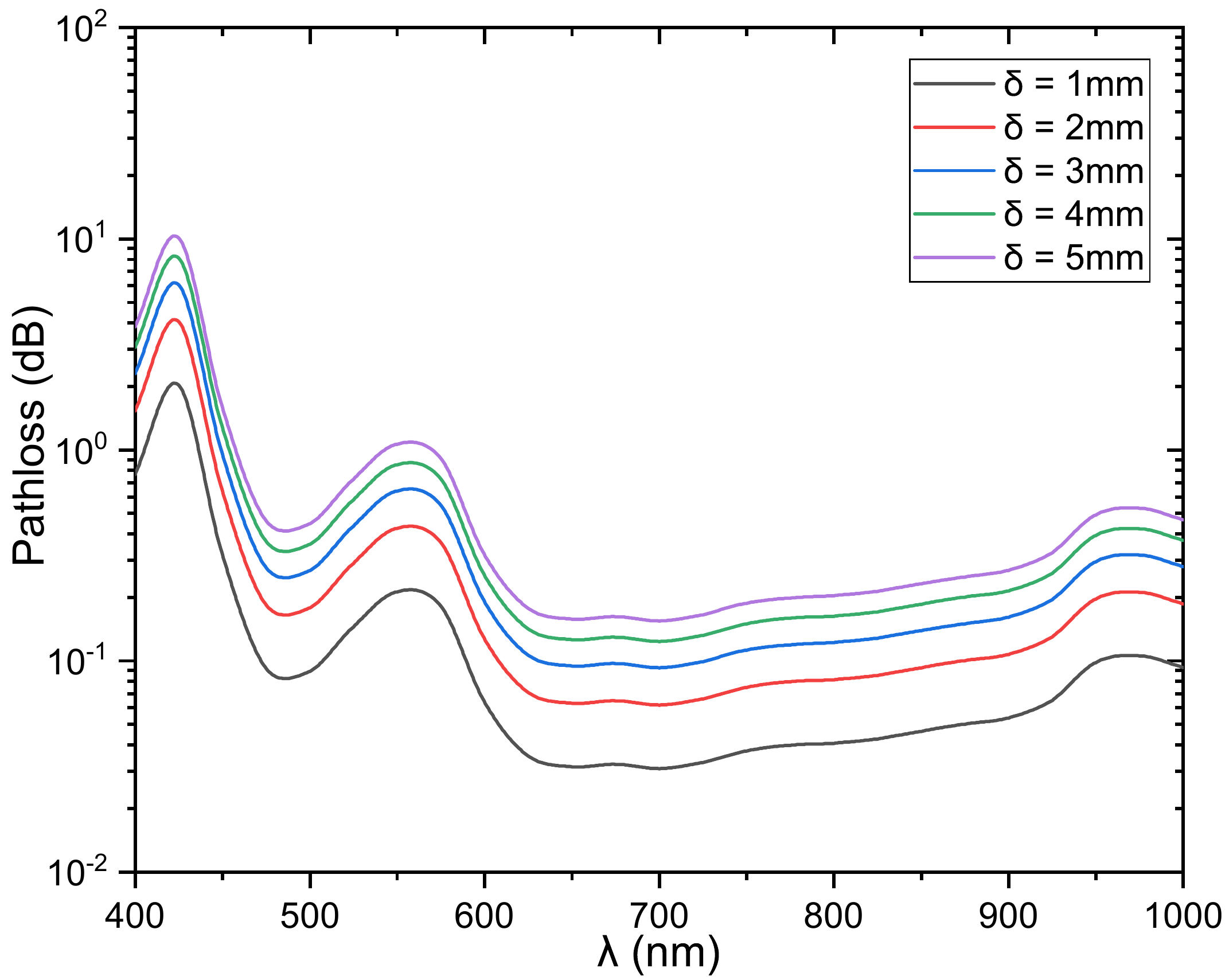}
		\caption{Pathloss due to bone absorption as a function of the transmission wavelength for different values of bone thickness.}\label{fig:bone_pathloss}
	\end{figure}
	In Fig.~\ref{fig:bone_pathloss}, the pathloss is depicted as a function of the transmission wavelength for various transmission distance values. Yet again, pathloss increases with tissue thickness, while its behavior with regard to $\lambda$ changes depends. The optimal $\lambda$ for transmission through bone tissue is $700\text{ }\mathrm{nm}$. For $\delta$ equal to $1$ and $2\text{ }\mathrm{mm}$, only one transmission window exists for $\lambda$ higher than $450\text{ }\mathrm{nm}$. On the contrary, for higher $\delta$ values, there are two transmission windows. For example, for $\delta = 5\text{ }\mathrm{mm}$, the two windows are $475 - 525\text{ }\mathrm{nm}$, and $600 - 950\text{ }\mathrm{nm}$.

	\section{Conclusion} \label{S:conclusion}
	In this paper, we first extracted analytical expressions for the absorption coefficient of the major generic tissue constituents based on published experimental measurements. These expressions enable the estimation of the absorption coefficient of each constituent at any given wavelength. Based on them, we formulate the mathematical framework for calculating the absorption coefficient of any generic tissue with regard to the transmission wavelength. Finally, we present the analytical results for the absorption coefficients of complex human tissues, compare them with experimental results from the open literature that prove the validity of the presented mathematical framework. Finally, we illustrate the pathloss as a function of the transmission wavelength for different complex tissues and tissue thickness, and provide insightful discussions.
	
	\section*{Aknowledgement}
	This research has been supported by the EU and Greek national funds through the Operational-Program Human-Resource-Development, Education and Lifelong Learning.
	
	\balance
	\bibliographystyle{IEEEtran}
	\bibliography{IEEEabrv,mybibfile}
	
\end{document}